\documentstyle[12pt]{article}

\begin{document}
\title{\textbf{Rotation and helicity as dynamo generators in dissipative and ideal plasma cosmologies}} \maketitle
{\sl \textbf{L.C. Garcia de Andrade}\newline
Departamento de F\'{\i}sica
Te\'orica-IF\newline
Universidade do Estado do Rio de Janeiro\\[-2mm]
Rua S\~ao Francisco Xavier, 524\\[-2mm]
Cep 20550-003, Maracan\~a, Rio de Janeiro, RJ, Brasil\\[-2mm]
Electronic mail address: garciluiz@gmail.com\\[-2mm]
\vspace{0.01cm} \newline{\bf Abstract} \paragraph*{}Recently Kleides et al [IJMPA \textbf{11}, 1697 (2008)] found a growing rate for magnetic fields in ideal plasma cosmologies by making use of general relativistic Friedmann model. This growth rate of $\frac{{\delta}B}{B}\sim{{(\frac{t}{t_{H}})}^{\frac{1}{4}}}$ indicates the presence of a slow dynamo in the universe. More recently Hasseein [Phys Plasmas (2009)] have also investigate Beltrami magnetic fields in plasma universe. Here general relativistic(GR) MHD dynamo equation, recently given by Clarkson and Marklund [Monthly Not Roy Astr Soc (2005)] is used to investigate the relation between collapsing of the isotropic universe and dynamo action in ideal and dissipative cosmologies. Dynamo action can be supported in these phases as long as the kinetic helicity overcomes universe diffusion effects. A cosmological Beltrami flow in 3D shows that helicities may act constructively on gravitational collapse and enhance dynamo action. A slow dynamo action is found in the static Einstein universe also filled with a Beltrami flow. A rotating, shear-free Bianchi type-IX universe, is obtained, by magnetically perturbing the Einstein static model inducing slow dynamos in the model. Magnetic field growth of $B\approx{t}$, which is stronger than Harrison estimate of $B\approx{t^{\frac{4}{5}}}$ is obtained. CMB limits on the expansion, global rotation and slow dynamos are given and a less slower dynamo than the one obtained by Kleides et al, is found with $\frac{{\delta}B}{B}\sim{|{\Theta}|t}$.\newline
\newpage
\section{Introduction}
Importance of the investigation of magnetic dynamos in the GR universe, has earlier been stressed by Zeldovich, Ruzmaikin and Sokoloff \cite{1}. More recently, Christensson and Hindmarsh \cite{2} have considered turbulent dynamos in Early Universe where a homogeneous cosmological model was used in connection with string theory. In this model no dynamo action was found and a decay of the magnetic was effective. More recently, Barrow and Tsagas (BT) \cite{3} in the form of a slow, decaying magnetic field in open Friedmann model. In this report one goes a small step further, and computes, based on CM MHD-GR dynamo equation, the existence of fast dynamo action \cite{4} in two distinct kinds of GR cosmologies. The first is the classical Einstein universe, where it is shown that the dynamo action is slow, for a magnetic curvature effect due to magnetic dissipation. Since in this case the shear, expansion and vorticity of cosmic fluid vanishes, the CM dynamo equation, leads to a simple version even in the presence of non-ideal plasmas. In this case, it is shown that, the fast dynamo action can be supported in this cosmology and only a slow dynamo is possible. This case represents a cosmological example of Vishik's anti-fast dynamo theorem. This problem has been investigated in detail recently by Garcia de Andrade \cite{5} in the case of dynamo plasmas \cite{6}. The magnetic field growth rate is computed in terms of eigenvalues spectrum of the dynamo operator. In the two and three-dimensions, compact Riemannian manifolds, the fast dynamo action has been obtained by Chicone et al \cite{7} in diffusive media. This has been done, following previous work by Arnold et al \cite{8}, in the case of uniform stretching of magnetic field by a steady flow in Riemannian space. Besides reproducing the decaying magnetic result of BT in the comoving coordinates, fast dynamo action is obtained, when the magnetic field growth rate of the Bianchi type-IX is positive. Effective growth rate is defined which includes the expansion of the universe. The stretching of universe cosmic plasma is obtained as a source for dynamo action. Stretching dynamos by plasma flows, have also been obtained by M Nu\~nez \cite{9}. An account of GR cosmological dynamos, is contained in Widrow \cite{10}. Section II of the paper, presents the Einstein static universe slow dynamo in diffusive media, while in section III, Bianchi type-IX dynamos are presented. In section IV it is shown that a slow dynamo in Bianchi type-IX model can be induced by magnetically perturbing the static Einstein cosmological model. Cosmic global rotation limits on slow dynamos is expected from cosmic background radiation (CMBR) data. Section V addresses the problem of similar kinetic effects on the existence of dynamo action in expanding phases of isotropic universes.
\section{Slow dynamos in Einstein static model}
The CM MHD-GR induction equations \cite{11} splitted into $(3+1)$-spacetimes can be recasted in the form used by dynamo theorists
\begin{equation}
[1+{\epsilon}(\frac{2}{3}{\Theta}+2{\vec{\omega}})]\dot{\textbf{B}}={\nabla}{\times}
(\textbf{v}{\times}\textbf{B})+{\epsilon}[{\Delta}\textbf{B}+2{\nabla}(\textbf{E}.\vec{\omega})-<\textbf{Ric},\textbf{B}>]+\vec{{\Gamma}}_{0}
\label{1}
\end{equation}
where ${\Delta}={\nabla}^{2}$ is the Laplacian, and
\begin{equation}
\vec{{\Gamma}}_{0}=(1+{\epsilon}\frac{2}{3}{\Theta})\textbf{B}
-{\epsilon}\textbf{F}-\textbf{G}+{\nabla}{\times}(\vec{\omega}{\times}\textbf{E})-
{\nabla}{\times}(\textbf{a}{\times}\textbf{E})
+{\epsilon}\vec{{\Gamma}}
\label{2}
\end{equation}
\begin{equation}
\textbf{F}=(\vec{\omega}{\times}\textbf{B}+\textbf{a}{\times}\textbf{E})\label{3}
\end{equation}
where in equation (\ref{1})
\begin{equation}
<\textbf{Ric},\textbf{B}>={R^{i}}_{j}B^{j}
\label{4}
\end{equation}
\begin{equation}
\textbf{G}=[\frac{2}{3}\textbf{E}{\times}{\nabla}{\Theta}-{\nabla}{\times}<{\sigma},\textbf{E}>
-{\nabla}{\times}(\vec{\omega}{\times}\textbf{E})]\label{5}
\end{equation}
Here greek letters shall denote space-time indices, while latin letters would represent three-dimensional space. Here also, $\textbf{Ric}$ represents the Ricci curvature tensor whose components are ${R^{i}}_{j}$. Here ${\textbf{E}}$, $\textbf{a}$,${\epsilon}$,${\Theta}$, ${\sigma}$ and ${\vec{\omega}}$ are respectively the, the electric field, the acceleration in three-space, the constant magnetic diffusion, or resistivity, and the respective Ehlers-Sachs optical cosmological parameters, expansion scalar, shear tensor and vorticity vector. The later shall be important in the next section when one shall address the rotational Goedel-like \cite{12} or Bianchi type-IX universe \cite{13}. one shall not reproduce here the long formula for ${\Gamma}(\textbf{a},{\Theta},{\sigma},\vec{\omega},\textbf{E})$ and in Einstein static universe in diffusive media it vanishes, since acceleration and Ehlers-Sachs optical parameters vanish and the electric field is here assumed to vanish as well, which is a reasonable assumption for most dynamo models \cite{1}. Since in the Einstein universe the Ricci tensor is given by the constant Riemann curvature contraction, called also Einstein manifold \cite{14}
\begin{equation}
R_{ij}={\Lambda}{g_{ij}}
\label{6}
\end{equation}
which Einstein latter recognised, as the cosmological constant, one may simplify the above CM MHD-GR dynamo equation as
\begin{equation}
[1+{\epsilon}(\frac{2}{3}{\Theta}+2{\vec{\omega}})]\dot{\textbf{B}}={\nabla}{\times}
(\textbf{v}{\times}\textbf{B})+{\epsilon}[{\Delta}\textbf{B}-<\textbf{Ric},\textbf{B}>]
\label{7}
\end{equation}
This equation can be further simplified, by assuming that
\begin{equation}
{\Delta}\textbf{B}=-curl(curl\textbf{B})
\label{8}
\end{equation}
and that the cosmic flow is a Beltrami flow, where
\begin{equation}
curl{\textbf{B}}={\nabla}{\times}\textbf{B}={\gamma}\textbf{B}
\label{9}
\end{equation}
which upon substitution into the above equation yields
\begin{equation}
{\Delta}\textbf{B}=-{\gamma}^{2}\textbf{B}
\label{10}
\end{equation}
Substitution of this last expression and the Ricci equation in Einstein manifold
yields
\begin{equation}
{\lambda}{\textbf{B}}=-{\epsilon}[{\gamma}^{2}-{\Lambda}]\textbf{B}
\label{11}
\end{equation}
since the static case leads to the vanishing of three-speed $\textbf{v}$. Here the dot represents partial time derivative, where one assumes that
\begin{equation}
{\textbf{B}}=\textbf{B}_{0}e^{{\lambda}t}
\label{12}
\end{equation}
Thus the eigenvalue spectrum of dynamo in the Einstein static universe , obtained from expression (\ref{19}) leads to
\begin{equation}
{\lambda}=-{\epsilon}[{\gamma}^{2}-{\Lambda}]
\label{13}
\end{equation}
Since the slow dynamo is given mathematically by the following limit
\begin{equation}
lim_{{\epsilon}\rightarrow{0}}{Re}({\lambda})=0
\label{14}
\end{equation}
where Re denotes the real part of the growth rate ${\lambda}$ which is in general complex, though here ${\lambda}\in{\textbf{R}}$, where $\textbf{R}$ if the field of real numbers. This is fulfilled by the Einstein universe, regardless whether or not, the cosmological constant is bound by the Beltrami constant ${\gamma}$. This is due to the fact that the above limit predominates over the fact if the limit above is positive or negative. Of course the fast dynamo shall be consider by the positivity of the growth rate ${\lambda}$. In the next section one shall discuss the fast dynamos in Bianchi type IX rotating cosmological model.
\newpage
\section{Bianchi type-IX cosmic fast dynamos in highly conducting plasmas}
In this section, one shall consider the Bianchi type-IX cosmology, given by the pseudo-Riemannian metric
\begin{equation}
ds^{2}=-dt^{2}+g_{ij}{\chi}^{i}{\chi}^{j}\label{15}
\end{equation}
where the differential basis one-form components ${\chi}^{i}$ are given by
\begin{equation}
{\chi}^{1}=-sin{x^{3}}dx^{1}+sinx^{1}cosx^{3}dx^{2}\label{16}
\end{equation}
\begin{equation}
{\chi}^{2}=cos{x^{3}}dx^{1}+sinx^{1}sinx^{3}dx^{2}\label{17}
\end{equation}
and
\begin{equation}
{\chi}^{3}=cosx^{1}dx^{2}+dx^{3}\label{18}
\end{equation}
In terms of the orthogonal tetrad frame, ${\omega}^{(i)}$ the pseudo-Riemannian metric maybe written as
\begin{equation}
ds^{2}=-dt^{2}+({\omega}^{(1)})^{2}+({\omega}^{(2)})^{2}+({\omega}^{(3)})^{2}\label{19}
\end{equation}
where
\begin{equation}
{\omega}^{(i)}={b^{i}}_{j}{\chi}^{(j)}\label{20}
\end{equation}
In terms of this "dreibein", ${b^{i}}_{j}$ the pseudo-Riemannian metric can be written as
\begin{equation}
g_{ij}={b_{i}}^{l}{b_{lj}}\label{21}
\end{equation}
The Bianchi type-IX cosmological models may expand or contract and have rotation according to the Ehlers-Sachs parameters, whose the only one that is missing here is the shear. Note that with the help of geodesic equation it is easy to obtain the values for ${\Theta}$ and rotation tensor of components ${\omega}_{ik}$ as
\begin{equation}
{\Theta}=\frac{l_{kj}v^{j}v^{k}}{v^{0}}-v_{0}l_{kk}\label{22}
\end{equation}
and
\begin{equation}
{\omega}_{0i}=-{\omega}_{i0}=-\frac{1}{2}\frac{v_{k}v_{l}{d^{k}}_{li}}{v_{0}}\label{23}
\end{equation}
\begin{equation}
{\omega}_{ij}=-{\omega}_{ji}=-\frac{1}{2}
v_{k}{d^{k}}_{ij}\label{24}
\end{equation}
where $v^{\alpha}$ is the four-spacetime velocity of the cosmic flow and
\begin{equation}
{d^{i}}_{jk}=b_{il}{\epsilon}_{lmn}{b_{mj}}^{-1}{b_{nk}}^{-1}\label{25}
\end{equation}
where
\begin{equation}
l_{ij}=k_{(ij)}=\frac{1}{2}(k_{ij}+k_{ji})\label{26}
\end{equation}
and
\begin{equation}
k_{ij}=\dot{b}_{il}{b_{lj}}^{-1}\label{27}
\end{equation}
By considering extremely highly conducting Bianchi type-IX universe, where the inverse of the magnetic Reynolds number $R_{m}$, ${\epsilon}={R_{m}}^{-1}$ vanishes, one simplifies the CM MHD dynamo equation leads to an effective magnetic field growth rate as
\begin{equation}
{\lambda}_{eff}=[{\lambda}+\frac{5}{3}{\Theta}+{{\Gamma}^{\alpha}}_{{\alpha}0}]
\label{28}
\end{equation}
where one has considered the comoving cosmic frame where $v^{0}=1$ and $\textbf{v}$ vanishes. This case is actually similar to the Friedman expanding universe. Since there is in this case a hypersurface orthogonality, vorticity vanishes, and only expansion survives. Here ${{\Gamma}^{\alpha}}_{{\beta}{\gamma}}$ is the Riemann-Christoffel symbol or Levi-Civita connection given by
\begin{equation}
{{\Gamma}^{\alpha}}_{{\beta}{\gamma}}=\frac{1}{2}g^{{\alpha}{\theta}}[-{{\partial}_{\theta}}g_{{\beta}{\gamma}}
+{{\partial}_{\gamma}}g_{{\beta}{\theta}}+
{\partial}_{\beta}g_{{\alpha}{\theta}}]\label{29}
\end{equation}
Computing the above component and the CM MHD GR dynamo equation
\begin{equation}
{\lambda}_{eff}{\textbf{B}}=\vec{\omega}{\times}\textbf{B}
\label{30}
\end{equation}
Since the RHS of this equation, vanishes due to the vanishing of vorticity, one obtains that vanishing of ${\lambda}_{eff}$ leads to
\begin{equation}
{\lambda}=-[\frac{5}{3}{\Theta}+{{\Gamma}^{\alpha}}_{{\alpha}0}]
\label{31}
\end{equation}
which yields finally the equation
\begin{equation}
{\lambda}=[\frac{5}{3}{l_{kk}}-{{\Gamma}^{\alpha}}_{{\alpha}0}]
\label{32}
\end{equation}
for the geodesic triad ${l^{k}}_{p}$. The existence of fast chaotic dynamos, depends upon the sign of the Riemann-Christoffel component. Thus, let us now compute this component in terms of the Bianchi type IX metric as
\begin{equation}
{{\Gamma}^{\alpha}}_{{\alpha}0}=k_{kk}
\label{33}
\end{equation}
Now this expression tell us that the growth rale can be given by
\begin{equation}
{\lambda}=-\frac{2}{3}{\Theta}
\label{34}
\end{equation}
To obtain this result one considers that the magnetic and velocity flow are solenoidal divergence-free vectors as
\begin{equation}
{\nabla}.{\textbf{B}}=0
\label{35}
\end{equation}
and
\begin{equation}
{\nabla}.{\textbf{v}}=0
\label{36}
\end{equation}
 This result agrees with BT and CM results that, when universe undergoes contraction (${\Theta}<0$) or stretching in the language of dynamo theorists, there is a stretching in the universe; the fast dynamo action is possible since ${\lambda}$ is positive. In what follows one shall prove that this result is possible to be obtained in the less trivial case of the rotating Bianchi type-IX, which is the main result of this report. Let us now consider then, the equation of the GR MHD-dynamo, in the case vorticity and expansion does not vanish is
\begin{equation}
[\lambda+\frac{5}{3}{\Theta}+(\textbf{a}.\textbf{v})+{{\Gamma}^{{\alpha}}}_{{\alpha}{\beta}}v^{\beta}]
{\textbf{B}}={\nabla}{\times}(\textbf{v}{\times}\textbf{B})+
{\vec{\omega}}{\times}\textbf{B}\label{37}
\end{equation}
Note that since the universe is assumed highly conducting ${\epsilon}$ vanishes and the curvature effects on the dynamo does not appear in the equation through the term $<\textbf{Ric},\textbf{B}>$ but actually through the Riemann-Christoffel symbol ${{\Gamma}^{{\alpha}}}_{{\alpha}{\beta}}$. This equation can be further simplified if one assumes that acceleration $\textbf{a}$ comes from the magnetic Lorentz force as
\begin{equation}
\textbf{a}=-\textbf{v}{\times}\textbf{B}\label{38}
\end{equation}
Due to simple vector analysis equations, this yields the following simplifications
\begin{equation}
\textbf{a}.\textbf{v}=-{\textbf{v}}.[\textbf{v}{\times}\textbf{B}]=0\label{39}
\end{equation}
and
\begin{equation}
\textbf{a}.\textbf{B}=-[\textbf{v}{\times}\textbf{B}].\textbf{B}=0\label{40}
\end{equation}
which reduces the dynamo equation (\ref{37}) to
\begin{equation}
{\nabla}{\times}(\textbf{v}{\times}\textbf{B})+
{\vec{\omega}}{\times}\textbf{B}=[\lambda+\frac{5}{3}{\Theta}+(\textbf{a}.\textbf{v})+{{\Gamma}^{{\alpha}}}_{{\alpha}{\beta}}v^{\beta}]
{\textbf{B}}\label{41}
\end{equation}
This last equation, can be recasted in dynamo operator form as
\begin{equation}
{\cal{L}}{\textbf{B}}={{\lambda}_{eff}}{\textbf{B}}\label{42}
\end{equation}
where the dynamo operator can be written as
\begin{equation}
{\cal{L}}={\nabla}{\times}\textbf{v}{\times}+
{\vec{\omega}}{\times}\label{43}
\end{equation}
while the effective eigenvalue growth rate is obtained as
\begin{equation}
{\lambda}_{eff}=[\lambda+\frac{5}{3}{\Theta}+{{\Gamma}^{{\alpha}}}_{{\alpha}{\beta}}v^{\beta}]
\label{44}
\end{equation}
Note that when one is on a syncronised comoving coordinate system where $v^{\beta}={{\delta}^{\beta}}_{0}$, expression (\ref{44}) reduces to the above equation for the Friedmann dynamo operator. Now let us compute the trace of the Riemann-Christoffel symbol, ${{\Gamma}^{{\alpha}}}_{{\alpha}{\beta}}$ contracted with the four-velocity $v^{\alpha}$ as
\begin{equation}
{{\Gamma}^{{\alpha}}}_{{\alpha}{\beta}}v^{\beta}={\nabla}_{\alpha}v^{\alpha}-
{\partial}_{\alpha}v^{\alpha}={\Theta}-{\nabla}.\textbf{v}={\Theta}
\label{45}
\end{equation}
Since locally the flow is incompressible, or divergence-free. Substitution of expression (\ref{45}) into (\ref{44}) yields
\begin{equation}
{\lambda}_{eff}=[\lambda+\frac{8}{3}{\Theta}]
\label{46}
\end{equation}
Thus in the case of zero-effective eigenvalue ${\lambda}_{eff}$, as a minimum effective eigenvalue, yields the following growth rate of the magnetic field as
\begin{equation}
\lambda=-\frac{8}{3}{\Theta}
\label{47}
\end{equation}
which shows that the fast dynamo action may exist in the Bianchi type-IX universe if the expansion ${\Theta}$ is negative. This, to a certain extent, explains why stellar colapsing objects which undergoes this gravitational contraction, may support fast dynamo action. The magnetic field in this case is
\begin{equation}
B^{i}={B^{i}}_{0}e^{-\frac{8}{3}{\Theta}t}
\label{48}
\end{equation}
which shows that when the Bianchi type-IX universe expands, (${\Theta}>0$) the magnetic field decay as in the Friedman case, the role of vorticity being so weak in cosmic dynamo here as is in the universe where expansion is much stronger than rotation as measured by COBE.
\section{Slow dynamo in Bianchi type-IX from magnetic perturbation of Einstein model}
In this section one stablishes a bridge between the last two sections by showing tha magnetic perturbation of the static Einstein model from a constant seed magnetic field, in the absence of shear, may lead to a slow dynamo in Bianchi type-IX model. Let us start by the CM equation in the case of highly conductive universe, which is a good approximation of our real world. The perturbation scheme is
\begin{equation}
\textbf{B}=\textbf{B}_{0}+\textbf{B}_{1}
\label{49}
\end{equation}
\begin{equation}
\textbf{v}=\textbf{v}_{0}+\textbf{v}_{1}
\label{50}
\end{equation}
\begin{equation}
{\Theta}={\Theta}_{0}+{\Theta}_{1}
\label{51}
\end{equation}
\begin{equation}
\vec{\omega}=\vec{\omega}_{0}+\vec{\omega}_{1}
\label{52}
\end{equation}
\begin{equation}
\textbf{a}=\textbf{a}_{0}+\textbf{a}_{1}
\label{53}
\end{equation}
Since the Einstein universe is static the following simplifications arise: All unperturbed quantities, but the magnetic seed field $\textbf{B}_{0}$, vanishes. Substitution of these results into the CM dynamo GR-MHD induction equation, yields
\begin{equation}
\dot{\textbf{B}}_{1}={\nabla}{\times}(\textbf{v}_{1}{\times}\textbf{B}_{0})-
\vec{\omega}_{1}{\times}\textbf{B}_{0}=(\textbf{v}_{1}.{\nabla})\textbf{B}_{0}-
(\textbf{B}_{0}.{\nabla})\textbf{v}_{1}-{\vec{\omega}_{1}}{\times}\textbf{B}_{0}\label{54}
\end{equation}
Note that this induction equation is linear in the seed field $\textbf{B}_{0}$, which allows us to simply integrate in time this equation, by assuming that the expansion is almost constant, or in the small acceleration phase. With the help of the equation
\begin{equation}
{\nabla}_{i}v^{k}={{\omega}_{i}}^{k}+{\Theta}{{\delta}_{i}}^{k}\label{55}
\end{equation}
The stretching term is given by
\begin{equation}
{{B_{0}}^{i}}.{\nabla}_{i}{v_{1}}^{k}={{B_{0}}^{i}}.[{{{\omega}_{1}}_{i}}^{k}+
{\Theta}_{1}{{\delta}_{i}}^{k}]\label{56}
\end{equation}
The other term is given by
\begin{equation}
[\vec{\omega}_{1}{\times}\textbf{B}_{0}]={\epsilon}_{ijk}{{\omega}_{1}}^{j}{{B}_{0}}^{k}
\label{57}
\end{equation}
These two terms cancell in equation
results into the CM dynamo GR-MHD induction equation, yields
\begin{equation}
\dot{\textbf{B}}_{1}={\nabla}{\times}(\textbf{v}_{1}{\times}\textbf{B}_{0})-
{\vec{\omega}_{1}}{\times}\textbf{B}_{0}\label{58}
\end{equation}
one obtains the equation
\begin{equation}
\dot{\textbf{B}}_{1}:={\delta}\textbf{B}=
-[\frac{1}{3}{\Theta}\textbf{B}_{0}+{\vec{\omega}_{1}}{\times}\textbf{B}_{0}]\label{59}
\end{equation}
which upon integration and Schwarz lemma reads
\begin{equation}
|\frac{{\delta}{B}}{B}|\le{(|\frac{1}{3}{\Theta}|+|\vec{\omega}{\times}\textbf{b}|)t}\label{60}
\end{equation}
where $\textbf{b}=\frac{\textbf{B}_{0}}{B_{0}}$ is the unit direction of seed magnetic field. Thus since the COBE data, given by
\begin{equation}
|\frac{{\delta}{B}}{B}|\le{10^{-3}}\label{61}
\end{equation}
provides a bound to the slow dynamo given by expression (\ref{60}). From dynamo theory this is expected, since due to dynamo quenching the dynamo magnetic field cannot grow without bounds. So in a certain sense one should say that the COBE data places limits not only on expansion and vorticity of the Bianchi type-IX model but also in the dynamo action. That is the reason, why the dynamo action is so rare in cosmological large scale dynamos, as pointed out previously by Barrow and Tsagas. Other global rotation limits from COBE in shear-free Bianchi type model has been proposed previously by Obukhov \cite{15} and Garcia de Andrade \cite{16} in Einstein-Cartan gravity, however, from pratical purposes, rotation in real universes, can be neglected in most situations. Nevertheless from the dynamo point of view rotation or twist, is fundamental as stretching, and is interesting to use models that contains rotation to check their influence on dynamo action. Here ${{B}}_{1}\approx{t}$. By using Zeldovich-Novikov gravitational perturbations in the expanding universe, Harrison \cite{17}. has actually found slower magnetic field growth as $B\approx{t^{\frac{4}{5}}}$. Taking $B_{1}={\delta}B$, expression (\ref{57}) becomes $|\frac{{\delta}B}{B}|=|{\Theta}|t$. This result is in agreement with the Friedman model result $|\frac{{\delta}B}{B}|=|\frac{\dot{H}}{H}|t\approx{10^{-3}}$.
\section{Dynamo action in expanding universes}
Based on their MHD dynamo equation, Clarkson and Marklund \cite{11} have shown that the cosmological gravitational collapse of the universe shall certainly leads to dynamo action while in expanding universes, the magnetic field would decay as shown in BT paper refereed above. In this section one shows that the universe expansion phases can actually support dynamo action as long as the mean electromotive force, given by
\begin{equation}
{\cal{E}}=\textbf{v}{\times}\textbf{B}={\alpha}\textbf{B}
\label{62}
\end{equation}
and ${\alpha}$ kinetic helicity effects overcomes universe expansion ${\Theta}$. This simple proof can be done by substitution of expression (\ref{62}) into the dynamo GR MHD formula (\ref{1}) by neglecting electric fields, yields
\begin{equation}
\dot{\textbf{B}}={\nabla}{\times}
({\alpha}\textbf{B})+{\epsilon}([{\Delta}\textbf{B}]-<\textbf{Ric},\textbf{B}>
+\vec{{\Gamma}}_{0})
\label{63}
\end{equation}
In the case of Beltrami flows, equations (\ref{9}) and (\ref{10}) yields
\begin{equation}
\dot{\textbf{B}}(1+\frac{5}{3}{\Theta}{\lambda})=
[({\alpha}-{\epsilon}){\gamma}-(\frac{2}{3}{\Theta}+\frac{1}{3}{\lambda}(E-3p+
4{\Lambda}+\frac{2}{3}{\Theta}^{2}))]
\textbf{B}\label{64}
\end{equation}
By assuming the diffusion is small and ${\cal{O}}({\epsilon}^{2})\rightarrow{0}$, one is able to obtain the constraints to helicity spectrum in order to obtain the dynamo action without necessarily to have collapse or ${\Theta}\le{0}$. Here the relation
\begin{equation}
R_{ij}=\frac{2h_{ij}}{3}{\lambda}(E+{\Lambda}-\frac{1}{3}{\Theta}^{2})\label{65}
\end{equation}
between the Ricci tensor, and energy density ${\epsilon}$, and the cosmological constant ${\Lambda}$ is used. Here $h_{ij}$ is the projective 3D metric. By assuming that a marginal dynamo is not a dynamo action, one may find a simple degenerate spectrum of expansion by assuming that $\dot{\textbf{B}}$ vanishes on the LHS of induction equation (\ref{65}) as
\begin{equation}
[({\alpha}-{\epsilon}){\gamma}-(\frac{2}{3}{\Theta}+\frac{1}{3}{\lambda}(E-3p+
4{\Lambda}+\frac{2}{3}{\Theta}^{2}))]=0
\label{66}
\end{equation}
This second order algebraic equation in the universe expansion ${\Theta}$ yields the real helicity spectrum as
\begin{equation}{\alpha}^{2}-2{\epsilon}{\alpha}+\frac{4{\epsilon}}{15}(E-3p+
4{\Lambda}+\frac{2}{3}{\Theta}^{2})=0
\label{67}
\end{equation}
which yields the following constraints for real spectrum
\begin{equation}
(R+2{\Lambda}+\frac{1}{3}{\Theta}^{2})\le{3p}
\label{68}
\end{equation}
which yields a highly pressured and early phase of the universe where helicity acts. Note that since the expansion appears squared in this condition does not mind if the universe is expanding or contracting for marginal dynamos. Note, however in this case, unless the Ricci scalar and cosmological constant be negative relation (\ref{68}) is violated. But this is exactly the gravitational collapse case. However, in the infinite conductivity case, a simple proof of the fact that even in expanding phases of the universe helicities could be strong enough to enhance dynamo action can be given. This can be done by considering that the above equations in this section reduce to
\begin{equation}
{\lambda}={\alpha}{\gamma}-\frac{1}{3}{\Theta}
\label{69}
\end{equation}
This shows that a fast dynamo action is possible even when the expansion of the universe $({\Theta}\ge{0})$ is taking place as long as the product of the kinetic and magnetic helicities are positive and strong enough to overcome expansion. This happens in the early phases in the universe. Of course this shall be stronger enhanced when the contraction phases taking place. That could explain why BT results where even in slow expansion a decay of magnetic fields is possible. Thus the universe expansion phases are more willing to possess decay magnetic fields than dynamo action.
\newpage
\section{Conclusions}
Thus one may conclude that by the analogous stretch-twist-fold Vainshtein-Zeldovich mechanism \cite{18} in cosmology given by expansion-rotation-curvature, one is able to obtaining amplification of the magnetic field, from a seed magnetic field by dynamo mechanism, producing at least a slow dynamo. Therefore the slow dynamos obtained here, from magnetically unstable Einstein static universe, endowed with a seed field, seems to be in agreement with previous ideas that the seed fields generated during phase transitions in the radiation era, have typical coherence lengths that, by its smallness,may destablize the dynamo. As also pointed out by Barrow et al \cite{19}, fields which survives inflationary epoch, are in general too weak to sustain a fast dynamo. It is interesting to note that, the smallness of the global rotation of the universe seems to lead to a slow dynamo, which eventually decays upon expansion. This eventually does not happen in the case one introduces mean field helicity into Clarkson Marklun MHD dynamo equations, as long helicity overcomes expansion.

\section{Acknowledgements}

 Several discussions with D Sokoloff, Yu N Obukhov, V Oguri and F Caruso Neto, are highly appreciated. Financial supports from UERJ and CNPq are also appreciated.
 

\begin{thebibliography}{15}
  \bibitem{1} Ya B Zeldovich, A Ruzmaikin and D D Sokoloff, Magnetic fields in the Astrophysics, (1980) G+B, New York.
  \bibitem{2} M Christensson and M Hindmarsh, Phys Rev D \textbf{60}(1999):063001. B Bassett,G Pollifrone, S Tsujikawa and F Viniegra, \textbf{Pre-heating: Cosmic dynamo?}, Phys Rev D \textbf{63}(2001): 103515.
  M Nu\~nez, J Phys \textbf{A},8903 (2003).
  \bibitem{3} J D Barrow and C Tsagas, Phys Rev D \textbf{77},107302 (2008).
  \bibitem{4} S Childress and A D Gilbert, Stretch-Twist and Fold: The fast dynamo, LNP ,(1995), Springer Verlag.
  \bibitem{5} L C Garcia de Andrade, The role of stretching and
  curvature in fast dynamo plasmas in Riemannian space, Phys Plasmas \textbf{15} (2008) in press.
  \bibitem{6} Z. Wang, V. Pariev, C. Barnes and D Barnes, Phys
  Plasmas \textbf{9},5 (2002).
  \bibitem{7} C. Chicone and Yu Latushkin, Evolution Semigroups in Dynamical systems and differential equations, American
  Mathematical Society, AMS-(1999). C. Chicone and Yu Latushkin and S. Montgomery-Smith,Comm. Math. Physics \textbf{173} 379 (1995). C. Chicone and Yu Latushkin, Proc of the American
  Mathematical Society 125, N. 11,3391 (1997).
  \bibitem{8} V. Arnold, Ya B. Zeldovich, A. Ruzmaikin and D.D. Sokoloff, JETP 81(1981),n. 6, 2052. V. Arnold, Ya B. Zeldovich, A. Ruzmaikin and D.D.
  Sokoloff, Doklady Akad. Nauka SSSR 266 (1982) n6, 1357.
  \bibitem{9} M Nu\~nez, J Phys \textbf{A},8903 (2003).
  \bibitem{10} L Widrow, Rev Mod Phys \textbf{74} (2002) 775.
  \bibitem{11} M Marklund and C Clarkson, Mont Not Roy Astron Soc (2005).
  \bibitem{12} K Goedel, Rev Mod Phys \textbf{21},(1949): 447.
  \bibitem{13} I Ciufolini and J A Wheeler, Gravitation and Inertia, (2004) Princeton series in Physics (1995).
  \bibitem{14} A Besse, Einstein manifolds, (2007) Springer-Verlag.
  \bibitem{15} Yu N Obukhov, On physical foundations and observational effects of cosmic rotation, in Colloquium on Cosmic Rotation (2000), edited by M Sheferner,T Chorobok and M Shefaat. Springer, Berlin.
  \bibitem{16} Garcia de Andrade, Ap Space Science (2004).
  \bibitem{17} E Harrison, Annual Rev Astronomy and Astrophysics, \textbf{11}, 155 (1973)
  \bibitem{18} S Vainshtein and Ya B Zeldovich, Sov Phys Usp \textbf{15}, 159 (1972).
  \bibitem{19} J D Barrow, R Martees and C Tsagas, Phys Reports \textbf{449}: 131 (2007).
  \end{thebibliography}
  \end{document}